\documentclass[12pt,preprint]{aastex}

\def\ltsima{$\; \buildrel < \over \sim \;$}
\def\gtsima{$\; \buildrel > \over \sim \;$}
\def\lsim{\lower.5ex\hbox{\ltsima}}
\def\gsim{\lower.5ex\hbox{\gtsima}}
\def\lapp{\ifmmode\stackrel{<}{_{\sim}}\else$\stackrel{<}{_{\sim}}$\fi}
\def\gapp{\ifmmode\stackrel{>}{_{\sim}}\else$\stackrel{<}{_{\sim}}$\fi}

\newdimen\minuswidth    
\setbox0=\hbox{$-$}
\minuswidth=\wd0
\catcode`@=\active
\def@{\kern\minuswidth}
\setbox0=\hbox{\rm0}
 
\shorttitle{The structural parameters of Terzan 5}
\shortauthors{Lanzoni et al.}
 
\begin{document} 
\title{New density profile and structural parameters of the complex
  stellar system Terzan 5\footnote{Based on observations collected at
    the European Southern Observatory, Cerro Paranal (during the
    Science Demonstration of MAD) and La Silla (under proposal
    278.D-5067), Chile. Also based on observations with the NASA/ESA
    {\it HST}, obtained at the Space Telescope Science Institute,
    which is operated by AURA, Inc., under NASA contract NAS5-26555.}
}

\author{
B. Lanzoni\altaffilmark{2}, 
F.R. Ferraro\altaffilmark{2},
E. Dalessandro\altaffilmark{2},
A. Mucciarelli\altaffilmark{2}, 
G. Beccari\altaffilmark{3},
P. Miocchi\altaffilmark{2}, 
M. Bellazzini\altaffilmark{4},
R.M. Rich\altaffilmark{5},
L. Origlia\altaffilmark{4},
E. Valenti\altaffilmark{6,7},
R.T. Rood\altaffilmark{8},
S.M. Ransom\altaffilmark{9}
}
\affil{\altaffilmark{2} Dipartimento di Astronomia, Universit\`a degli Studi
di Bologna, via Ranzani 1, I--40127 Bologna, Italy}
\affil{\altaffilmark{3}
ESA, Space Science Department, Keplerlaan 1, 2200 AG Noordwijk, Netherlands}   
\affil{\altaffilmark{4}
INAF- Osservatorio Astronomico di Bologna, Via Ranzani, 1, 40127 Bologna, Italy}
\affil{\altaffilmark{5}
Department of Physics and Astronomy, 
Math-Sciences 8979, UCLA, Los Angeles, CA 90095-1562,USA}
\affil{\altaffilmark{6}P. Universidad Catolica de Chile, 
Departamento de Astronomia y Astrofisica, 
Casilla 306, Santiago 22, Chile}
\affil{\altaffilmark{7} European Southern Observatory,
 Alonso de Cordova 3107, Vitacura, Santiago, Chile}
\affil{\altaffilmark{8}Astronomy Department, University of Virginia, 
P.O. Box 400325, Charlottesville, VA, 22904,USA}
\affil{\altaffilmark{9}
National Radio Astronomy Observatory, Charlottesville, VA 22903, USA} 
\date{12 Mai, 2010}

\begin{abstract}
Terzan~5 is a globular cluster-like stellar system in the Galactic
Bulge which has been recently found to harbor two stellar populations
with different iron content and probably different ages
\citep{fe09}. This discovery suggests that Terzan~5 may be the relic
of a primordial building block which contributed to the formation of
the Galactic Bulge. Here we present a re-determination of the
structural parameters (center of gravity, density and surface
brightness profiles, total luminosity and mass) of Terzan~5, as
obtained from the combination of high-resolution ({\it ESO}-MAD and
{\it HST} ACS-WFC) and wide-field ({\it ESO}-WFI) observations. We
find that Terzan~5 is significantly less concentrated and more massive
than previously thought. Still it has the largest collision rate of
any stellar aggregate in the Galaxy.  We discuss the impact of these
findings on the exceptional population of millisecond pulsars harbored
in this stellar system.
\end{abstract}
 
\keywords{Globular clusters: individual (Terzan 5); stars: evolution}

\section{INTRODUCTION}
Terzan~5 is commonly catalogued as a globular cluster (GC) located in
the inner Bulge of our Galaxy.  It is difficult to observe because it
is heavily reddened, with an average color excess $E(B-V)=2.38$
(Barbuy et al. 1998; Valenti et al. 2007). Not only is the reddening
large, but it strongly depends on the line of sight (differential
reddening; see Ortolani et al. 1996, Valenti et al. 2007).  Terzan~5
has an exceptionally large population of millisecond pulsars
(MSPs). Indeed, the 33 MSPs detected so-far in Terzan~5 amount to
about $25\%$ of the entire sample of known MSPs in Galactic GCs
(Ransom et al. 2005; see the updated list at {\tt
  http://www.naic.edu/pfreire/GCpsr.html})

\begin{figure}[!hp]
\begin{center}
\includegraphics[scale=0.9]{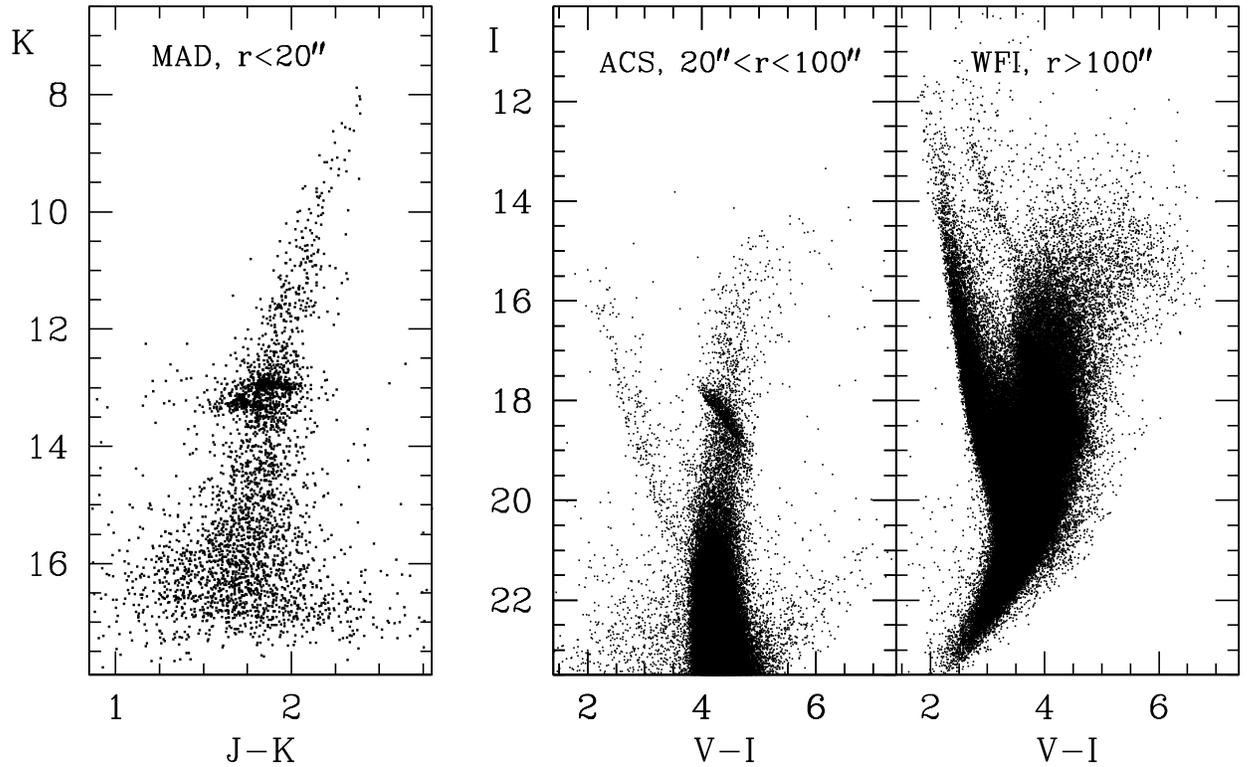}
\caption{Color-magnitude diagrams for the three datasets used in the
  paper and for different radial regions of the cluster (see
  labels). The double HB discovered by F09 is clearly visible in the
  MAD near-infrared CMD, while it is more difficult to distinguish in
  the ACS optical plane (but see Supplementary Figure 1 in F09). The
  WFI sample is dominated by field star contamination.}
\label{fig:cmd}
\end{center}
\end{figure}

As part of a project (Ferraro et al. 2001, 2003; Cocozza et al. 2008)
aimed at studying the properties of stellar populations harboring
MSPs, we obtained a set of high-resolution images of Terzan~5 in the
$K$ and $J$ bands using the multi-conjugate adaptive optics (AO)
demonstrator MAD \citep{marchetti07} temporally installed at the
European Southern Observatory (ESO) Very Large Telescope (VLT).  The
($K,~J-K$) color-magnitude diagram (CMD) obtained from these
observations led to the discovery of two well-defined red horizontal
branch (HB) clumps, clearly separated in luminosity ($\delta K\sim
0.3$) and color (see Ferraro et al. 2009, hereafter F09; also see
Figure \ref{fig:cmd}). A prompt spectroscopic follow-up demonstrated
that the two populations have the same radial velocity (hence they
belong to the same stellar systems) and their metal content is
different: ${\rm [Fe/H]} \simeq -0.2$ and ${\rm [Fe/H]}\simeq +0.3$
for the fainter and the brighter group, respectively. These findings
and the comparison with theoretical stellar isochrones confirm the
existence of two distinct stellar populations in Terzan~5 and suggest
that they possibly have been generated by two bursts of star formation
with a time separation of a few ($\approx 6$) Gyrs.  While the age gap
can be reduced by invoking a difference in the helium content of the
two populations \citep{dantona10}, the iron enrichment and the spatial
segregation of the brightest clump, together with the extraordinary
amount of MSPs found in Terzan 5, indicate that this system probably
experienced a particularly troubled formation and evolutionary history
(see Sect. 5).  Terzan~5 is the first a GC-like system in the Galactic
Bulge found to have a spread in the iron content and it could be the
relic of one of the building blocks that contributed to the formation
of the Bulge.  Indeed the discovery might represent the observational
evidence that even the innermost part of galactic spheroids form (at
least partially) by the accretion/merging of small, previously formed
and internally evolved stellar systems \citep[e.g.,][]{immeli04}.

In this paper we present the accurate re-determination of Terzan~5
structural parameters (surface density and surface brightness
profiles, total luminosity, collision rate, etc.), obtained from a
combination of high-resolution and wide-field observational
data. These parameters provide basic information for a deeper
understanding of the origin and the evolution of this puzzling system.

\section{THE DATA}
The photometric dataset used in the present work consists of
high-resolution and wide-field images obtained, respectively, with MAD
at the ESO-VLT and the Advanced Camera for Survey (ACS) on board the
Hubble Space Telescope (HST), and with the Wide Field Imager (WFI) at
the 2.2m ESO-MPI telescope.

{\it 1. The MAD-dataset} consists of a set of short (2-minute long),
AO-corrected exposures secured through the $K$ and $J$ filters in
August 2008, as part of a MAD science demonstration project
(P.I. Ferraro).  From all these images, we chose and analysed the
highest quality ones: the full width at half maximum (FWHM) measured
for the selected $K$ and $J$ images is $0.1\arcsec$ and $0.24\arcsec$,
respectively, just slighly larger than the diffraction limit. More
importantly the FWHM is extremely stable over the entire
($1\arcmin\times 1\arcmin$) MAD field of view (FOV), fully
demonstrating the potentiality of the multi-conjugate AO correction.

{\it 2. The HST-dataset} consists of deep ACS Wide Field Camera (WFC)
images obtained through filters F606W (a broad $V$) and F814W ($I$),
with total exposure times of 340 s and 360 s, respectively
(Prop. 9799, P.I. Rich).

{\it 3. The Wide field-dataset} consists of multiple $V$ and $I$
images ($4\times 120$ s exposures each) obtained with the ESO-WFI at
La Silla (Chile) and retrieved from the ESO Science Archive
(Prop. 278.D-5067(A); P.I. Bassa). The WFI is a mosaic of 8-CCD chips
which combines wide-field (FOV =$33\arcmin\times 34\arcmin$) and
reasonably high-resolution capabilities (pixel size of $\sim
0.24\arcsec$).  The core of the cluster is roughly centered on CCD 7
and the images allow to sample the entire cluster extension (see
Sect. \ref{sec:prof}).

The point spread function (PSF) for each image has been modelled on
several bright and isolated stars, by using the DAOPHOTII/PSF routine
\citep{ste87}.  Then PSF-fitting photometry of the MAD dataset has
been perfomed independently on the $K$ and $J$ best images, using
DAOPHOTII/ALLSTAR.  We then used DAOPHOTII/MONTAGE2 procedure to
produce a stacked master frame combining the optical images in all the
available filters (F606W and F814W for the ACS observations, and $V$
and $I$ for the WFI dataset). This method allowed us to obtain, for
each optical dataset, a single high S/N image, cleaned from specific
detector defects (i.e. hot pixels) and other spurious sources like
cosmic rays etc.  A master star list has been searched at a $4\sigma$
detection limit on this reference frame and it has then been used as
input for ALLFRAME \citep{ste94}, which simultaneously determines the
brightness of the stars in all the frames, while enforcing one set of
centroids and one transformation between all the images. Finally, the
magnitudes obtained for each star have been normalized to a reference
frame and averaged together, and the photometric error was derived
from the standard deviation of the repeated measures.

The star positions in the WFI sample were placed on the absolute
astrometric system by using more than twenty thousand stars in common
with the new astrometric Two Micron All Sky Survey (2MASS)
catalogue\footnote{Available at {\tt
    http://irsa.ipac.caltech.edu}.}. Then the $\sim 6000$ stars in
common between the WFI and the ACS datasets have been used as
secondary astrometric standards for placing the ACS sample on the
absolute astrometric system, and the same has been done for the $\sim
9000$ stars in common between the ACS and the MAD datasets, following
the procedure described, e.g., in \citet{lan07}. The final astrometric
accuracy of all the samples is of the order of $\sim 0.2\arcsec$ both
in right ascension ($\alpha$) and in declination ($\delta$). The
near-infrared instrumental magnitudes have been reported to the 2MASS
photometric system by using the stars in common with the catalog of
\citet{valenti07}. The optical magnitudes have been transformed and
homogenized by using a sample of stars in common with Ortolani et
al. (1996).

The CMDs obtained for the three datasets are shown in
Figure\,\ref{fig:cmd}. The two HB clumps discussed by F09 are clearly
visible in the MAD dataset and partially distinguishable in the ACS
sample, while the field contamination (mainly from the Galactic Bulge
and Disk) is progressively more important at increasing distance ($r$)
from the cluster center. The optical CMD clearly shows how Terzan~5 is
strongly affected by differential reddening. We are currently
constructing a detailed differential reddening map in the direction of
this stellar system (Mucciarelli et al. 2010, in preparation), and
preliminary results suggest that the reddening variation can be as
high as $\delta\, E(B-V)=0.6$.

\section{CENTER OF GRAVITY}
F09 determined the center of gravity ($C_{\rm grav}$) of Terzan~5 by
using the absolute positions of individual stars detected in the MAD
sample.  As a check, here we have recomputed $C_{\rm grav}$ by
exploiting the ACS dataset, with a cut in magnitude $I=21$ needed to
avoid spurious effects due to incompleteness in the very inner regions
of the cluster.  We found that, within the uncertainty ($\sim
0.5\arcsec$ in both $\alpha$ and $\delta$), the two determinations are
coincident, and we therefore confirm that gravity center of Terzan~5
is located at (F09): $\alpha_{\rm J2000} = 17^{\rm h}\, 48^{\rm m}\,
4.85^{\rm s}$, $\delta_{J2000} = -24\arcdeg\,46\arcmin\,
44.6\arcsec$. This is $\sim 1\arcsec$ north-west from the photometric
centre quoted by \citet{harris96}, and $\sim 1\arcsec$ north-east from
that of \citet{cohn02}.

To further check the reliability of our determination we also used a
different approach. This relies on the algorithm defined by
\citet{casehut85} to calculate the "density center" of a stellar
system (for more details see Miocchi et al. 2010, in preparation).  As
a first step, for all the $N$ stars located within a circle of a given
radius, centered at an initially guessed position, the local (surface)
density $\rho_i$ around the position of the $i$th star is evaluated as
the inverse of its squared distance from the sixth nearest star. Then
the cluster center is computed as the density-weighted average of the
$N$ star positions. The evaluation is repeated by centering the circle
at the last found points and the iteration stops when the distance
between the new and the previous determinaton is smaller than a given
value (usually $0.01\arcsec$). The density center thus computed agrees
with the value quoted above well within the uncertainties.  

\begin{figure}[!hp]
\begin{center}
\includegraphics[scale=0.9]{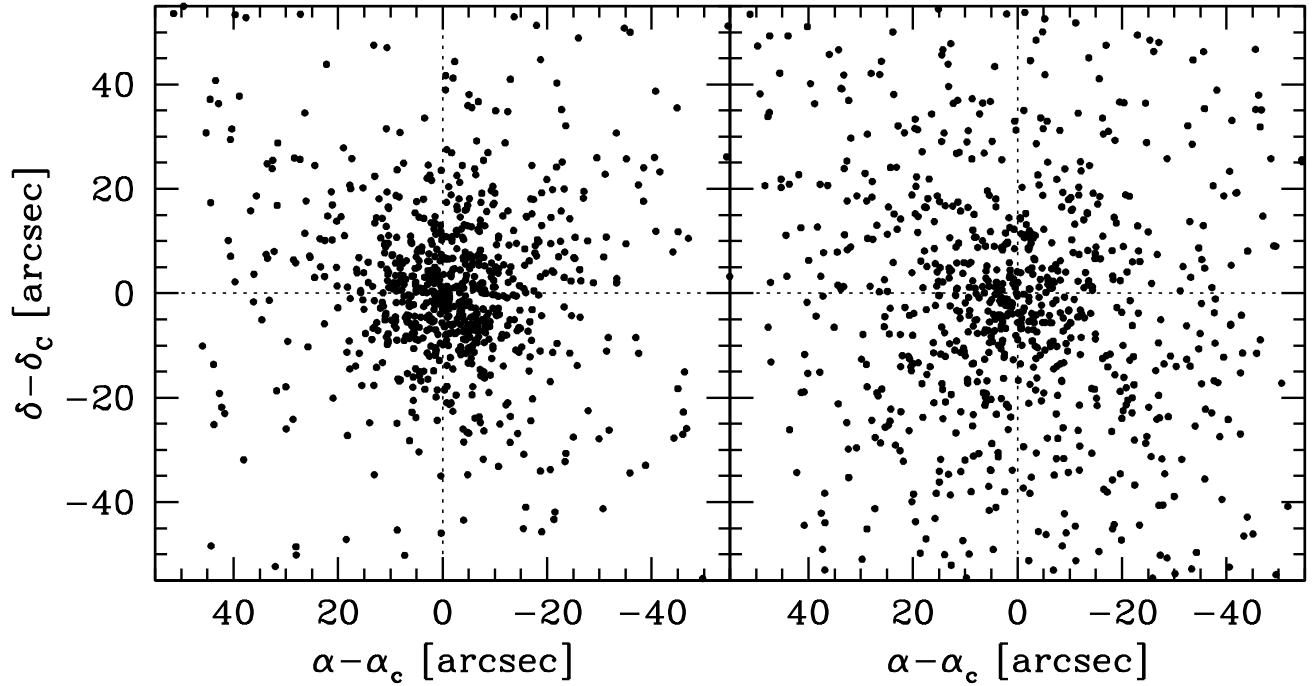}
\caption{Map of the bright ({\it left panel}) and faint ({\it right
    panel}) HB populations selected from the ACS dataset on the basis
  of the star position in the ($I,~V-I$) CMD. The star coordinates
  ($\alpha$ and $\delta$) are plotted with respect to those of the
  cluster gravity centre ($\alpha_{\rm c}$ and $\delta_{\rm c}$).  }
\label{fig:hbmap}
\end{center}
\end{figure}

Finally, we also checked whether the two HB populations share the same
gravity centre. To this purpose we have selected them from the ACS
dataset, on the basis of the star position in the ($I,~V-I$) CMD (see
Supplementary Figure 1 in F09).  The projected spatial distribution of
the two selected samples is plotted in Figure \ref{fig:hbmap} and
clearly shows that the bright HB population is more centrally
concentrated than the faint one, in agreement with what found by F09
from the MAD dataset.  The barycentres of the two HB populations seem
to be different, with the faint HB centre being located $\approx
3\arcsec$ south-east from that of the bright HB stars, which almost
coincides with the cluster gravity centre quoted above. We stress
however that, while the optical selection allowed us to improve the
statistics and increase the radial coverage of the samples at most, it
does not guarantee a proper separation of the two groups of
stars. Hence, before confirming such a finding it is necessary to
perform a more robust and clean selection of the two HB populations,
based on (presently not available) near-infrared data covering a much
larger area than the MAD FOV.

\section{PROJECTED DENSITY AND SURFACE BRIGHTNESS PROFILES}
\label{sec:prof}
\begin{figure}[!hp]
\begin{center}
\includegraphics[scale=0.7]{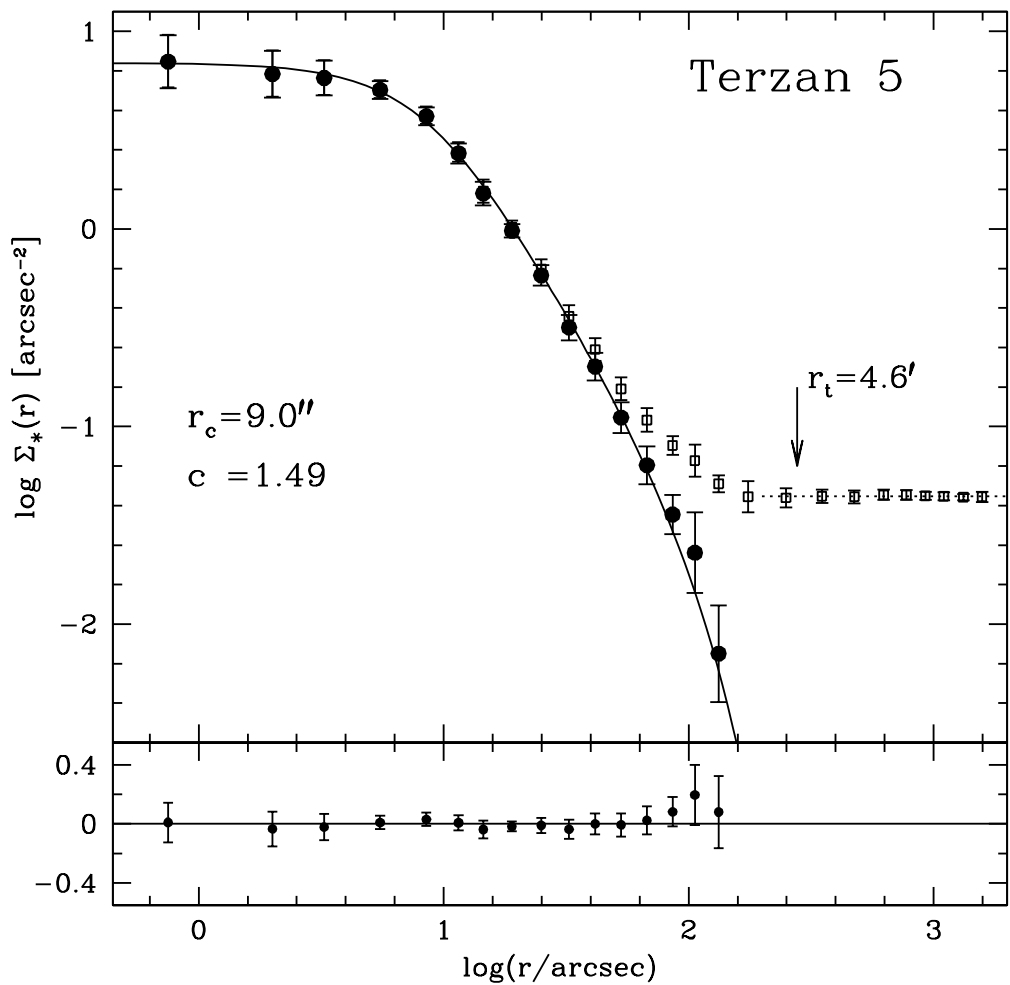}
\caption{Newly determined star density profile of Terzan~5, obtained
  from resolved star counts in the combined photometric dataset.
  Empty squares represent the observed profile, while solid dots are
  obtained after subtraction of the field background density (marked
  with the dotted line). The best-fit single-mass King model is shown
  as a solid line and the corresponding structural parameters (core
  radius $r_c$, concentration $c$ and tidal radius $r_t$) are labelled
  in the figure. The lower panel shows the residuals between the
  observations and the fitted profile at each radial coordinate.}
\label{fig:prof}
\end{center}
\end{figure}

We have determined the projected density profile of Terzan~5 using
direct star counts on the available datasets, which cover the entire
radial extent of the cluster.  For the innermost part of the profile
we have exploited the high-resolution (MAD and ACS) datasets, while
for $r>100\arcsec$ we have used the WFI sample and complementary data
from the 2MASS survey, thus covering radial distances out to
$r=1700\arcsec$.  To avoid incompleteness biases, different limiting
magnitudes have been adopted for the four datasets: $K=13$, $I=20$,
$I=19$, and $K=12.5$ for the MAD, ACS, WFI, and 2MASS samples,
respectively. By also adopting the colour cuts $(V-I)>3.4$ and
$(J-K)>1.3$ we have excluded from the analysis the contribution of the
stars belonging to the Galactic disk main sequence. With these limits
we have computed the four portions of the density profile
corresponding to each dataset.\footnote{While different limits have
  been (necessarily) adopted for the four datasets, the mass of the
  sampled stars is roughly constant (comparable to the that of the
  main sequence turn off stars). Hence any mass segregation effect on
  the four different portions of the profile is expected to be
  negligible. This is further confirmed by the agreeement found
  between the number density and the surface brightness profiles (see
  below).} In total more than 50,000 stars were used to construct the
star density profile.  Following the procedure described in
\citet{fe99} we have divided the samples in concentric annuli centered
on $C_{\rm grav}$, and each annulus has been split into an adequate
number of sub-sectors. The number of stars lying within each
sub-sector was counted, and the star surface density was obtained by
dividing these values by the corresponding sub-sector areas.  The
stellar density in each annulus was then obtained as the average of
the sub-sector densities, and the standard deviation was adopted as
the uncertainty.  Then, the radial annuli in common between the
different samples have been used to shift and join the various
portions of the profile. The overall projected density profile thus
obtained is shown in Figure\,\ref{fig:prof} (empty squares), with the
abscissae corresponding to the mid-point of each radial bin.  The
outermost ($r\gsim 175\arcsec$) measures from the WFI and 2MASS
samples have an almost constant value, and their average has been used
to estimate the Galactic Bulge and Disk contamination level. The
subtraction of this background yields the profile shown in the figure
as filled dots.  The derived density profile is well fit all over its
radial extension by an isotropic, single-mass King model
\citep{king66} with core radius $r_c=9.0\arcsec$, half-mass radius
$r_h=31\arcsec$, tidal radius radius $r_t=277\arcsec=4.6\arcmin$, and
intermediate concentration ($c=1.49$).  While the size of the core
radius is consistent with the most recent determination
($r_c=7.9\arcsec$; Cohn et al. 2002), the concentration is
significantly smaller than that ($c\approx2$) suggested by those
authors, and the ratio between the core and the half-mass radius is a
factor of two larger in our case.

\begin{center}
\begin{figure}[!p]
\includegraphics[scale=0.7]{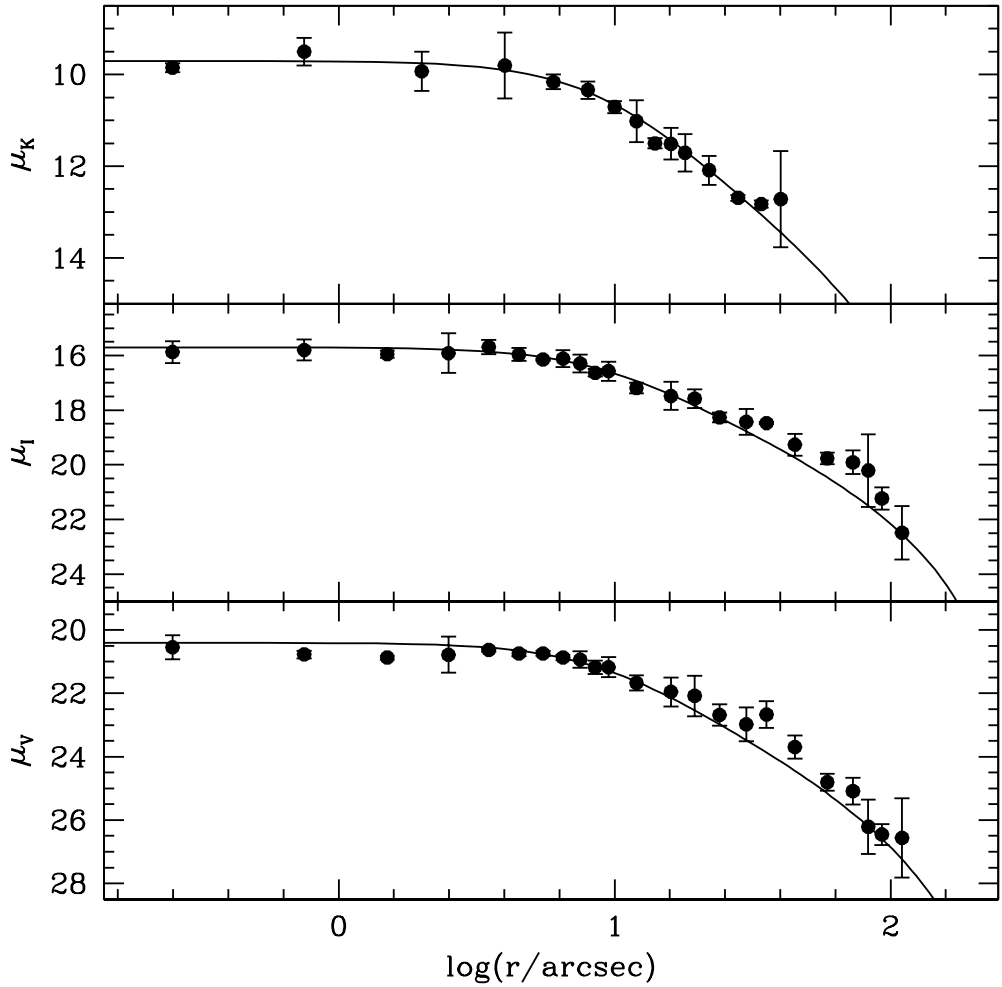}
\caption{Surface brightness profile of Terzan~5 in three different
  filters, after background subtraction: $K$ band (upper panel),
  obtained from the MAD dataset, $I$ and $V$ bands (middle and lower
  panels), from the ACS sample. In each panel the solid line
  corresponds to the best-fit King model shown in
  Fig.\,\ref{fig:prof}.}
\label{fig:mu}
\end{figure}
\end{center}

Exploiting the exceptional quality of the available datasets, we have
also computed the surface brightness (SB) profiles by aperture
photometry on the MAD and ACS images. The SB values were computed as
the sum of the photon counts in each pixel, divided by the sampled
area in any given radial annulus. The counts have been converted to a
magnitude scale and then calibrated using a relation derived by
performing aperture photometry on a number of high S/N isolated stars.
The resulting SB profile, obtained for the inner $\sim 100\arcsec$
from the center, after proper subtraction of the background level and
of a few artefacts due to saturated stars, is shown in three different
filters in Figure \ref{fig:mu}.  These profiles are well fit by the
same King model derived from the projected density distribution, thus
confirming that the samples of resolved stars used above are not
affected by radial variations of the completeness and are properly
selected.  The values of the central SB measured in the three
photometric bands are listed in Table 1, together with all the
relevant parameters derived for Terzan~5.

\section{DISCUSSION}
The star density and SB profiles can be used to derive the integrated
luminosity of the cluster. From the best-fit King model we estimate
that the percentage of cluster light within regions of radius
$r=15\arcsec$, $18\arcsec$, and $20\arcsec$ are roughly 30\%, 36\% and
40\%, respectively.  Using aperture photometry on the MAD images, we
obtain integrated-light values of $K(r<15\arcsec) = 3.44$,
$K(r<18\arcsec) = 3.3$ and $K(r<20\arcsec) = 3.2$ mag, respectively.
Adopting the color excess $E(B-V)=2.38$, the distance modulus
$(m-M)_0=13.87$ (Valenti et al. 2007, corresponding to a distance of
$d=5.9\pm0.5$ kpc) and the bolometric correction $BC_K = 2.4$
appropriate for a population of intrinsic color $(J-K)_0 = 0.8$ (see
Montegriffo et al. 1998), we estimate that the corresponding
bolometric luminosity in the considered regions is: $L_{\rm bol} (r <
15\arcsec) = 3 \times 10^5 L_\odot$, $L_{\rm bol} (r < 18\arcsec) =
3.4 \times 10^5 L_\odot$ and $L_{\rm bol} (r < 20\arcsec) = 3.7 \times
10^5 L_\odot$.  Considering the fraction of light sampled in each
region we find that the total luminosity of the system is $L_{\rm bol}
= 9.5\pm0.3 \times 10^5 L_\odot$.

An independent estimate of the total luminosity of the stellar system
can be derived from its stellar population, by using a simple relation
\citep{fuel86} linking the number of stars ($N_j$) observed in a given
post-main sequence evolutionary stage $j$ and the luminosity of the
entire parent cluster ($L_T$):
\begin{equation}
N_j = B \times t_j \times L_T,
\end{equation}
where $B$ is the specific evolutionary flux (for intermediate/old
stellar populations $B = 2 \times 10^{-11}\,{\rm stars\, yr}^{-1}
L_\odot^{-1}$) and $t_j$ is the duration of the evolutionary stage.
The number of HB stars counted in the two clumps by F09 is quite
large: a total of about 1300 (with 800 and 500 belonging to the faint
and the bright HB clumps, respectively). This population is comparable
to (or even larger than) that observed in the largest Galactic GCs,
like 47 Tucanae \citep{bec_47tuc} and NGC~6388 \citep{ema_6388}, and
suggests that the overall size of Terzan 5 (in terms of luminosity and
mass) is comparable to that of these systems. For a quantitative
estimate, we insert the observed number of HB stars in the above
relation and adopt $t_{\rm HB} = 10^8$ yr. This provides a luminosity
of $4 \times 10^5 L_\odot$ and $2.5 \times 10^5 L_\odot$ for the two
parent populations, and a total luminosity of $6.5 \times 10^5
L_{\odot}$ for the entire stellar system.  This estimate, which is
distance and reddening independent, is quite consistent with the
previous one, thus confirming that Terzan~5 has a considerable total
luminosity (hereafter we adopt the average value $L_{\rm bol} = 8 \times
10^5 L_\odot$), significantly higher than previously thought. By
comparison, adopting the values of distance and reddening quoted above
and a bolometric correction $L_{\rm bol}\simeq 1.4\,L_V$, the total
bolometric luminosity corresponding to the integrated magnitude
($V_t=13.85$) quoted by \citet{harris96} would be only $L_{\rm bol}\simeq
10^5 L_\odot$.  The discrepancy is most probably due to the strong
(differential) reddening affecting the system, especially in the
optical bands. This effect is greatly reduced for our new estimate,
since it is based on the observed $K$-band integrated magnitude and
the number of HB stars.  By assuming a mass-to-light ratio
$M/L_{\rm bol}=3$ \citep[e.g.,][]{maraston98}, the total stellar mass of
this system is $M_{\rm T} \simeq 2 \times 10^6 M_\odot$.

\citet{verhut87} first suggested that the collision rate of Terzan~5
is the highest among the Galactic GCs. We can now re-compute this
quantity by adopting the newly determined parameters. Following
\citet{verhut87}, the collisional parameter ($\Gamma$) for a
King-model or virialized system can be computed as: $\Gamma \propto
\rho_0 \times r_c^{0.5}$, where $\rho_0$ is the central mass density.
By using the values obtained above and equation (7) of
\citet{djorg93}, we find that the collision parameter of Terzan~5 is
between 5 and 10 times higher than that of Liller 1 and of other
massive clusters for which structural parameters have been recently
re-determined \citep[NGC6388, NGC6266, 47Tuc;][respectively]{ema_6388,
  bec_6266, mapelli06}. Hence we confirm that, even with the new
structural parameters (suggesting a lower concentration and a larger
mass than previously thought), Terzan~5 still has the largest known
collision rate of any stellar aggregate in the Galaxy.

The co-existence of two stellar populations with different iron
content (and probably ages) suggests that the original mass of
Terzan~5 was significantly larger in the past than observed today,
large enough to retain the iron-enriched gas that, otherwise, would
have been ejected out from the system by the violent supernova (SN)
explosions.  Indeed, the smallest systems with solid evidences of a
spread in the iron content (and ages) are significantly more massive
than GCs: the dwarf spheroidal satellites of the Milky Way typically
have masses of $\sim 10^7 M_\odot$ \citep[Strigari et al. 2008; see
  also][]{battaglia08} and, following recent chemo-dynamical models
well reproducing the observations, their initial masses amounted to a
few $10^8 M_\odot$ \citep{revaz09}. While a lower limit of $\sim 10^7
M_\odot$ for the proto-Terzan 5 could also be hazarded following
\citet{baumgardt08}, more detailed and extensive simulations are
needed to firmly determine the smallest total mass necessary to retain
the SN ejecta.

The exceptionally high metallicity regime of the two stellar
populations found in Terzan~5 also suggests a quite efficient
enrichment process, that could have a relevant role in the origin of
its population of MSPs. In particular, both the iron and the
[$\alpha/$Fe] abundance ratios measured in Terzan~5 \citep[][Rich et
  al. 2010, in preparation]{origlia04} show a remarkable similarity
with those of the Bulge stars.  This strongly suggests that these two
structures shared the same star formation and chemical enrichment
processes.  The many observations of Bulge stars \citep[e.g.,][and
  references therein]{melendez08,origlia08,ryde09} indicate that they
are all characterized by an old age, a high (close to solar) average
metallicity [Fe/H], and an [$\alpha/$Fe] ratio which is enhanced (due
to SNII enrichment) up to a metallicity [Fe/H]$\simeq 0$. These
constraints suggest a scenario where the dominant stellar population
of the Bulge formed early (thus explaining the old
age\footnote{\scriptsize{Additional episodes of star formation mainly
    confined in the innermost ($\sim 100$ pc) region could eventually
    explain the presence of younger stars
    \citep[e.g.,][]{blum03,figer04}}.}), rapidly and with high
efficiency (from a gas mainly enriched by SNII, thus explaining the
[$\alpha/$Fe] enhancement up to high iron
contents\footnote{\scriptsize{The [$\alpha/$Fe]--[Fe/H] relation shows
    a down-turn at a value of [Fe/H] which depends on the star
    formation rate: the higher the latter, the higher the metallicity
    at which the down-turn occurs. Such a value is [Fe/H]$\simeq -1$
    in the Old Halo/Disk, while it is significantly higher
    ([Fe/H]$\simeq 0$) in the Bulge, testifying a much higher star
    formation rate in this dense environment.}}).  Also chemical
evolution models \citep[e.g.,][]{ballero07,mcwilliam08} indicate that
the abundance patterns observed in the Bulge require a quite high star
formation efficiency and an initial mass function flatter than that in
the solar neighbourhood, thus to rapidly enrich the gas up to about
solar metallicity through an exceptionally large amount of SNII
explosions. The assumption of a similar scenario for Terzan~5 would
naturally explain its extraordinary population of MSPs, since the
expected high number of SNII would produce a large population of
neutron stars, most of which would have been retained by the deep
potential well of the massive proto-Terzan~5 system.  Then the high
collision rate could have favoured the formation of binary systems
containing neutron stars and promoted the re-cycling process that
finally generated the large population of MSPs now observed in
Terzan~5. If such a scenario is correct, many more MSPs still wait to
be discovered in this system \citep[see also][]{ransom05}, the 33
known objects probably being just the tip of the iceberg. Future
deeper pulsar searches of Terzan~5, perhaps with larger telescopes
such as the Square Kilometer Array, will shed additional light on the
nature of this system.

\acknowledgements This research was supported by the Agenzia Spaziale
Italiana (under contract ASI-INAF I/016/07/0), by the Istituto
Nazionale di Astrofisica (INAF, under contract PRIN-INAF2008) and by
the Ministero dell'Istruzione, dell'Universit\`a e della Ricerca. RTR
is partially supported by STScI grant.  RMR is supported by
AST-0709479 and GO-9799 from STScI. This research has made use of the
ESO/ST-ECF Science Archive facility which is a joint collaboration of
the European Southern Observatory and the Space Telescope - European
Coordinating Facility.

\begin{deluxetable}{ll}
\tablecolumns{2}
\footnotesize
\tablecaption{Structural parameters for Terzan~5}
%
\tablewidth{9cm}
\startdata \\
\hline  
\hline
   Center of gravity & $\alpha_{\rm J2000} = 17^{\rm h}\, 48^{\rm
  m}\, 4.85^{\rm s}$\\
~ & $\delta_{J2000} = -24\arcdeg\,46\arcmin\, 44.6\arcsec$  \\
   Reddening$^\dagger$ & $E(B-V) = 2.38\pm 0.055$   \\
   Distance$^\dagger$  & $d=5.9\pm 0.5$ kpc\\
   Core radius & $r_c = 9\arcsec = 0.26$ pc \\
   Concentration & $c=1.49$ \\
   Total luminosity & $L_{\rm bol}\simeq 8\times 10^5\, L_\odot$ \\
   Total mass & $M_{\rm T}\simeq 2 \times 10^6 M_\odot$\\
   Central mass density & $\rho_0\simeq 4.1\times 10^6\, M_\odot/$pc$^3$ \\
   Central $K$-band SB & $\mu_K(0) = 9.85$ mag/arcsec$^2$\\
   Central $I$-band SB & $\mu_I(0) = 15.87$ mag/arcsec$^2$\\
   Central $V$-band SB & $\mu_V(0) = 20.54$ mag/arcsec$^2$\\
\hline 
\enddata
\tablecomments{$^\dagger$from \citet{valenti07}}
\label{tab:counts}
\end{deluxetable}

\begin{thebibliography}{}
 
\bibitem[Ballero et al.(2007)]{ballero07} Ballero, S.~K., Matteucci,
  F., Origlia, L., \& Rich, R.~M.\ 2007, \aap, 467, 123

\bibitem[Barbuy et al. (1998)]{barbuy98} Barbuy, B., Bica, E., \&
  Ortolani, S.\ 1998, A\&A, 333, 117

\bibitem[Battaglia et al.(2008)]{battaglia08} Battaglia, G., Helmi,
  A., Tolstoy, E., Irwin, M., Hill, V., \& Jablonka, P.\ 2008, \apjl,
  681, L13

\bibitem[Baumgardt et al.(2008)]{baumgardt08} Baumgardt, H., Kroupa,
  P., \& Parmentier, G.\ 2008, \mnras, 384, 1231

\bibitem[Beccari et al. (2006a)]{bec_47tuc} Beccari, G., Ferraro,
  F.R., Lanzoni, B., \& Bellazzini \ 2006a, \apjl, 652, 121

\bibitem[Beccari et al. (2006b)]{bec_6266} Beccari, G., Ferraro,
  F.~R., Possenti, A., Valenti, E., Origlia, L., \& Rood, R.~T.,
  2006b, AJ, 131, 2551

\bibitem[Blum et al.(2003)]{blum03} Blum, R.~D., Ram{\'{\i}}rez,
  S.~V., Sellgren, K., \& Olsen, K.\ 2003, \apj, 597, 323

\bibitem[Casertano \& Hut(1985)]{casehut85} Casertano, S., \& Hut,
  P.\ 1985, \apj, 298, 80

\bibitem[Cocozza et al.(2008)]{cocoz08} Cocozza, G., Ferraro, F.~R.,
  Possenti, A., Beccari, G., Lanzoni, B., Ransom, S., Rood, R.~T., \&
  D'Amico, N.\ 2008, \apjl, 679, L105

\bibitem[Cohn et al. (2002)]{cohn02} Cohn, H. N., Lugger, P. H.,
  Grindlay, J. E., \& Edmonds, P. D.\ 2002, \apj, 571, 818

\bibitem[Dalessandro et al. (2008)]{ema_6388} Dalessandro E., Lanzoni,
  B., Ferraro, F.R., Rood, R.T., Milone, A., Piotto, G., \& Valenti E.
  et al.\ 2008, \apj, 6677, 1069

\bibitem[D'Antona et al.(2010)]{dantona10} D'Antona, F., Ventura, P.,
  Caloi, V., D'Ercole, A., Vesperini, E., Carini, R., \& Di
  Criscienzo, M. \ 2010, ApJL in press (arXiv:1004.3426)

\bibitem[Djorgovski (1993)]{djorg93} Djorgovski, S.\ 1993, ASPC, 50, 373
 
\bibitem[Ferraro et al.(1999)]{fe99} Ferraro F. R., Messineo M., Fusi
  Pecci F., De Palo M. A., Straniero O., Chieffi A.,\& Limongi
  M.\ 1999, \aj, 118, 1738

\bibitem[Ferraro et al. (2001)]{fe01} Ferraro, F. R., 
Possenti, A., D'Amico, N., \& Sabbi, E.\ 2001, \apjl, 561, L93 

\bibitem[Ferraro et al. (2003)]{fe03} Ferraro, F. R., Possenti, A.,
  Sabbi, E., \& D'Amico, N.\ 2003, \apjl, 596, L211

\bibitem[Ferraro et al. (2009)]{fe09} Ferraro, F.~R., et al.\ 2009,
  \nat, 462, 483 (F09)
 
\bibitem[Figer et al.(2004)]{figer04} Figer, D.~F., Rich, R.~M., Kim,
  S.~S., Morris, M., \& Serabyn, E.\ 2004, \apj, 601, 319

\bibitem[Harris (1996)]{harris96} Harris, W.E. 1996, AJ, 112, 1487

\bibitem[Immeli et al.(2004)]{immeli04} Immeli, A., Samland, M.,
  Gerhard, O., \& Westera, P.\ 2004, \aap, 413, 547

\bibitem[King (1966)]{king66} King I.R. 1966, AJ, 71, 64

\bibitem[Lanzoni et al. (2007)]{lan07} Lanzoni, B., Dalessandro, E.,
  Ferraro, F.~R., Mancini, C., Beccari, G., Rood, R.~T., Mapelli, M.,
  \& Sigurdsson, S. \ 2007, ApJ, 663, 1040

\bibitem[Mapelli et al.(2006)]{mapelli06} Mapelli, M., Sigurdsson, S., Ferraro,
F.~R., Colpi, M., Possenti, A., \& Lanzoni, B.\ 2006, \mnras, 373, 361

\bibitem[Maraston(1998)]{maraston98} Maraston, C.\ 1998, \mnras, 300,
  872

\bibitem[Marchetti et al. (2007)]{marchetti07} Marchetti, E., et
  al. \ 2007, The Messenger, 129, 8

\bibitem[McWilliam et al.(2008)]{mcwilliam08} McWilliam, A.,
  Matteucci, F., Ballero, S., Rich, R.~M., Fulbright, J.~P., \&
  Cescutti, G.\ 2008, \aj, 136, 367

\bibitem[Mel{\'e}ndez et al. (2008)]{melendez08} Mel{\'e}ndez, J., et
  al.\ 2008, \aap, 484, L21

\bibitem[Montegriffo et al. (1998)]{mont98} Montegriffo, P., Ferraro,
  F. R., Origlia, O., \& Fusi Pecci, F.\ 1998, MNRAS, 297, 872

\bibitem[Origlia \& Rich (2004)]{origlia04} Origlia, L., \& Rich,
  R.~M.\ 2004, \aj, 127, 3422

\bibitem[Origlia et al.(2008)]{origlia08} Origlia, L., Valenti, E., \&
  Rich, R.~M.\ 2008, \mnras, 388, 1419

\bibitem[Ortolani et al. (1996)]{ortolani96} Ortolani, S., Barbuy, B.,
  \& Bica, E. 1996, A\&A, 308, 733
 
\bibitem[Ransom et al.(2005)]{ransom05} Ransom, S.~M., Hessels,
  J.~W.~T., Stairs, I.~H., Freire, P.~C.~C., Camilo, F., Kaspi, V.~M.,
  \& Kaplan, D.~L.\ 2005, Science, 307, 892

\bibitem[Renzini \& Buzzoni (1986)]{fuel86} Renzini, A. \& Buzzoni,
  A., 1986, Spectral evolution of galaxies; Dordrecht, D. Reidel
  Publishing Co., 195-231

\bibitem[Revaz et al.(2009)]{revaz09} Revaz, Y., et al.\ 2009, \aap,
  501, 189

\bibitem[Ryde et al. (2009)]{ryde09} Ryde, N., Edvardsson, B.,
  Gustafsson, B., Eriksson, K., K{\"a}ufl, H.~U., Siebenmorgen, R., \&
  Smette, A.\ 2009, \aap, 496, 701

\bibitem[Stetson (1987)]{ste87} Stetson, P.~B.\ 1987, \pasp, 99, 191

\bibitem[Stetson (1994)]{ste94} Stetson, P.~B.\ 1994, \pasp, 106, 250

\bibitem[Strigari et al.(2008)]{strigari08} Strigari, L.~E., Bullock,
  J.~S., Kaplinghat, M., Simon, J.~D., Geha, M., Willman, B., \&
  Walker, M.~G.\ 2008, \nat, 454, 1096

\bibitem[Valenti et al. (2007)]{valenti07} Valenti, E., Ferraro,
  F. R., \& Origlia, L., \ 2007, \aj, 133, 1287

\bibitem[Verbunt \& Hut (1987)]{verhut87} Verbunt, F., \& Hut,
  P. 1987, in Proc. IAU Symp. 125, The Origin and Evolution of Neutron
  Stars, ed. D. J. Helfand \& J.-H. Huang (Dordrecht: Reidel), 187

\end{thebibliography}
\end{document}